%% file: main.tex
\documentclass[twocolumn,amsmath, amssymb, superscriptaddress, aps, prd, 10pt, longbibliography]{revtex4-2}

\input{_preamble}

\begin{document}

\title{Joint Electromagnetic and Gravitational Wave Inference of Binary Neutron Star Merger GW170817 Using Forward-Modeling Ejecta Predictions}

\author{Marko~Risti\'c\,\orcidlink{0000-0001-7042-4472}}
\thanks{ISTI Fellow}
\email{mristic@lanl.gov}
\affiliation{\CTA}
\affiliation{\TD}
\author{Richard~O'Shaughnessy\,\orcidlink{0000-0001-5832-8517}}
\affiliation{\RIT}
\author{Kate~Wagner\,\orcidlink{0000-0002-7255-4251}}
\affiliation{\RIT}
\author{Christopher~J.~Fontes\,\orcidlink{0000-0003-1087-2964}}
\affiliation{\CTA}
\affiliation{\XCP}
\author{Chris~L.~Fryer\,\orcidlink{0000-0003-2624-0056}}
\affiliation{\CTA}
\affiliation{\CCSS}
\affiliation{\UA}
\affiliation{\NM}
\affiliation{\GWU}
\author{Oleg~Korobkin\,\orcidlink{0000-0003-4156-5342}}
\affiliation{\CTA}
\affiliation{\TD}
\author{Matthew R. Mumpower\,\orcidlink{0000-0002-9950-9688}}
\affiliation{\CTA}
\affiliation{\TD}
\author{Ryan~T.~Wollaeger\,\orcidlink{0000-0003-3265-4079}}
\affiliation{\CTA}
\affiliation{\CCSS}

\date{\today}

\begin{abstract}
 We reassess the capacity for multimessenger inference of AT2017gfo/GW170817 using both kilonova and gravitational wave emission within the context of a recent simulation-based surrogate model for kilonova emission. Independent of the inclusion of gravitational wave
 observations, comparisons between observations that incorporate our kilonova
 model favor a narrow range of ejecta properties, even when allowing for a wide range of systematic uncertainties in our
 modeling approach.  Conversely,  we find that astrophysical conclusions about the neutron
 star itself, including its mass and radius, depend strongly on assumptions about how much material is ejected from the
 neutron star. Looking forward, our analysis highlights the importance of systematic uncertainty in general, the need
 for better modeling of neutron star merger mass ejection from first principles, and warns against uncontextualized applications of ejecta predictions using fits to numerical relativity simulations.
 
\end{abstract}

\maketitle

\section{Introduction}
\label{sec:intro}

The detection of gravitational waves (GWs) emitted by \editedtwo{the} binary neutron star (BNS) merger GW170817 \cite{LIGO-GW170817-bns, LIGO-GW170817-mma} and, subsequently, electromagnetic (EM) radiation from its associated \edited{optical/infrared} counterpart AT2017gfo \cite{2017PASA...34...69A,2017Natur.551...64A,2017Sci...358.1556C,2017ApJ...848L..17C,Diaz17,2017Sci...358.1570D,Evans_2017,2017SciBu..62.1433H, 2017Sci...358.1559K, 2017ApJ...850L...1L,2017Natur.551...67P,2017Natur.551...75S,tanvir17, 2017Natur.551...71T,2017PASJ...69..101U,2017ApJ...848L..24V, 2018ApJ...852L..30P} introduced exciting prospects for studying the universe using multimessenger astronomy.
The GWs emitted by BNS mergers carry information about the pre-merger binary, such as the masses, spins, and tidal deformabilities of the merging neutron stars.
Likewise, the EM radiation from the \edited{optical/infrared} counterpart, or ``kilonova" \cite{1998ApJ...507L..59L, 2010MNRAS.406.2650M}, conveys details about the material that was ejected during and after the merger \cite{2019ARNPS..69...41S}, tying directly to the fate of the merger remnant \cite{2019MNRAS.483.1912P, 2021ApJ...908..152M, 2024ApJ...961L..26C}, the dense matter equation of state (EOS) \cite{LIGO-GW170817-EOS, Nicholl21}, and rapid neutron capture process (\emph{r}-process) nucleosynthesis \cite{2017Natur.551...80K, 2017ARNPS..67..253T, 2019EPJA...55..203S}.

Following the GW170817/AT2017gfo detection, recent studies have performed numerous numerical relativity simulations of \editedtwo{BNS} mergers \cite{2011PhRvD..83l4008H, 2017PhRvD..95d4045D, 2018ApJ...869..130R, 2019ApJ...876L..31K, 2013ApJ...773...78B, 2021ApJ...906...98N, 2020PhRvD.101d4053V, 2016PhRvD..93l4046S, 2013PhRvD..87b4001H, 2018ApJ...869..130R, 2016CQGra..33r4002L, 2019ApJ...876L..31K, 2019EPJA...55..124P, 2017PhRvD..95d4045D, 2019ApJ...886L..30N, 2020MNRAS.497.1488B, 2015PhRvD..91f4059S}.
Using these simulations, certain studies identified fitting formulae that connect \editedtwo{BNS merger} ejecta properties to the parameters of the binary that are easily extractable from the GW signal \cite{2020PhRvD.101j3002K, 2020Sci...370.1450D, 2022CQGra..39a5008N}.
\editedtwo{Mapping these ejecta properties to those relevant to kilonova model ejecta is a crucial ingredient in enabling multimessenger inferences using both GW and EM information.}
These fits \editedtwo{and ejecta mappings} have been used in previous multimessenger analyses to place stronger constraints on unknown quantities like the EOS \cite{2020Sci...370.1450D, 2022ApJ...926..196H, 2023NatCo..14.8352P}, particularly in the case of GW190425, a BNS/black-hole-neutron-star merger whose ejected mass had to be inferred using one of these fits due to the lack of an electromagnetic counterpart \cite{2021ApJ...922..269R}.
\editedtwo{It is important to note that the BNS merger ejecta to kilonova ejecta mapping is highly non-trivial, and an accurate one-to-one mapping still eludes the field. 
Numerical relativity models of BNS mergers are still computationally expensive, with recent record-breaking simulations only going out to $\sim 1.5$ s post-merger \citep{2025PhRvL.134u1407H}. The intrinsic variability in the BNS parameter space, in addition to kilonova time and length scales being many orders of magnitude larger than the longest numerical BNS models, necessitates extrapolation or projection in the mapping of BNS merger ejecta to kilonova model ejecta.
}

Previous work in the literature has identified that these fits to numerical relativity simulations exhibit variable validity in different
regions of binary parameter space \cite{2022MNRAS.516.4760C, 2023PhRvD.107f3028H}.  Moreover, the underlying simulations used to build
these fits only cover a very limited range of possible \editedtwo{BNS} merger configurations.  Nonetheless, as we will demonstrate in
this work, large-scale multimessenger inference calculations must employ these fits far outside the regions in which
they have been calibrated \edited{to recover reasonable binary parameters}. 

The need for using these fits outside of their safely-calibrated region has not been emphasized
in the many previous studies or frameworks that have employed them. \edited{Several (though not all) previous
  multimessenger inference studies \citep{2018MNRAS.480.3871C, 2019MNRAS.489L..91C, 2020Sci...370.1450D,
    2023NatCo..14.8352P} present joint EOS inference with systematically incomplete ejecta fit models, by marginalizing over the ejecta velocity \emph{for \editedtwo{all} kilonova \editedtwo{ejecta} components}, leaving ejecta velocity inference as a free parameter, unconstrained by any modeling. In particular, these studies neglect to verify whether the velocities required by the kilonova models to reproduce observed light curves are even qualitatively consistent with the velocity expectations from the ejecta fits. Furthermore, these analyses were carried out by
reweighting GW samples during joint inference, thus carrying a convenience prior that hides the effects that arise when performing simultaneous joint inference as explored in this study.}

Motivated by a recently developed surrogate model for kilonova emission that provides the most constraining ejecta \edited{posteriors} to date \cite{Peng24} \edited{(which do not necessarily encapsulate the full breadth of kilonova modeling systematics)}, in conjunction with previous studies examining the agreement between ejecta fits (e.g. \cite{2023PhRvD.107f3028H}), in this study we examine how these ejecta fits inform multimessenger inference of the binary parameters.
In this work, we assume a specific mapping between these ejecta types given its prior use in previous works and focus solely on the possible impact of the fitting formulae.
We caution readers that our assumption of ejecta mapping is by no means complete or one-to-one, but is a convention readily assumed in the field.
We emphasize that our kilonova model is also by no means complete in the sense that many systematic uncertainties are still omitted. Recent studies have identified multiple additional sources of systematic uncertainty along the merger-to-kilonova modeling chain \cite{2023mgm..conf.1391F, 2024ApJ...970..173M, 2024ApJ...975..213B}.
Nevertheless, our surrogate model's ability to tightly constrain ejecta parameters enables us to investigate the systematic uncertainties that propagate into Bayesian inference analyses that utilize these fits.
\edited{Our approach contrasts previous studies which assume kilonova systematic uncertainties anywhere between 1 magnitude \cite{2017CQGra..34j5014D, 2017ApJ...849...12C, 2018MNRAS.480.3871C} and 5 magnitudes \cite{2024A&A...689A..51B}; whereas the ejecta fit systematics have historically been washed out by the kilonova systematics, we take our surrogate model's constrained posteriors for granted and directly examine the systematics associated \emph{primarily} with the ejecta fits.}
Our intent with this study is to highlight the systematic uncertainties that are propagated when using these fits for more informed use in future studies.

The paper is structured as follows. Section~\ref{sec:methods} outlines our methodology, including details about the numerical relativity ejecta fits, our kilonova surrogate model and the simulations on which it is based, and our Bayesian inference approach. In Section~\ref{sec:results}, we discuss the results of our analysis, specifically focusing on the cases where the fraction of the accretion disk unbound as ejecta is treated as a free versus fixed parameter. In Section~\ref{sec:discussion}, we discuss our results and the implications of our findings. Finally, we conclude the paper in Section~\ref{sec:conclusion}.

\section{Methodology}
\label{sec:methods}

At a high level, our methodology proceeds as follows: we generate iterative grids of samples across our binary parameter space over which the likelihood is evaluated. For the GW data, we perform a coordinate transformation of our binary parameters to $\mathcal{M}_c$ (chirp mass), $q$ (mass ratio), $\tilde{\Lambda}$ (effective tidal deformability), and $\chi_{\rm{eff}}$ (effective spin) and calculate the GW likelihood through comparison to the GW170817 waveform. For the EM data, we calculate the ejecta properties from the binary parameter samples using the ejecta fits, create light curves corresponding to these ejecta properties, and calculate the EM likelihood through comparison to the AT2017gfo light curves. We continue generating iterative sample grids, informed by the prior iteration's likelihood, until we reach convergence ($\mathcal{O}(1000)$ effective samples).

\subsection{Kilonova simulations and surrogate model}
\label{sec:kn_sim_setup}

We assume a two-component kilonova model consisting of lanthanide-rich, equatorial dynamical ejecta and lanthanide-poor, axial wind ejecta as described in \cite{kilonova-lanl-WollaegerNewGrid2020, 2021ApJ...910..116K} and motivated by numerical simulations \citep{2019ARNPS..69...41S, just23}. Each component is assumed to be homologously expanding and parameterized by a mass and velocity such that ($m_{\rm{d}}$, $v_{\rm{d}}$) and ($m_{\rm{w}}$, $v_{\rm{w}}$) describe the dynamical and wind components' masses and averaged velocities, respectively. The morphology for the dynamical component is an equatorially centered torus, while the wind component is represented by an axially-centered peanut component; Figure~1 of \cite{kilonova-lanl-WollaegerNewGrid2020} displays the torus-peanut, or ``TP," schematic corresponding to the morphologies employed in this work; see Ref. \cite{2021ApJ...910..116K} for a detailed definition. 

The lanthanide-rich dynamical ejecta stems from $r$-process nucleosynthesis in the neutron-rich ejected material, characterized by a low electron fraction ($Y_{\rm{e}} \equiv n_{\rm{p}}/(n_{\rm{p}} + n_{\rm{n}})$) of $Y_e = 0.04$, with elements reaching the third $r$-process peak ($A \sim 195$). \edited{Our dynamical ejecta explores the nucleosynthesis resulting from the full $r$-process; however, recent studies have shown that the dynamical ejecta can be less neutron rich when considering neutrino interactions (see, e.g., Ref. \cite{2022MNRAS.510.2804K}).} The wind ejecta originates from higher $Y_e = 0.27$ that encapsulates elements between the first ($A \sim 80$) and second ($A \sim 130$) $r$-process peaks. The detailed breakdown of the elements in each component can be found in Table~2 of Ref.~\cite{kilonova-lanl-WollaegerNewGrid2020}.

To generate the simulations on which the surrogate model discussed in this work is trained, we use \texttt{SuperNu} \citep{SuperNu}, a Monte Carlo code for simulation of time-dependent radiation transport with matter in local thermodynamic equilibrium. The simulated kilonova spectra $F_{\lambda, \rm sim}$ assume the aforementioned two-component model. Both components are assumed to have fixed composition and morphology for the duration of each simulation. \texttt{SuperNu} uses radioactive power sources calculated from decaying the $r$-process composition from the \texttt{WinNet} nuclear reaction network \citep{2012ApJ...750L..22W,Korobkin_2012,Reichert2023a,Reichert2023b}. These radioactive heating contributions are also weighted by thermalization efficiencies introduced in Ref.~\cite{Barnes_2016} (see Ref.~\cite{Wollaeger2018} for a detailed description of the adopted nuclear heating). We use detailed opacity calculations via the tabulated, binned opacities generated with the Los Alamos suite of atomic physics codes \citep{2015JPhB...48n4014F,2020MNRAS.493.4143F,nist_lanl_opacity_database}. In the database that we use, the tabulated, binned opacities are not calculated for all elements; therefore, we produce opacities for representative proxy elements by combining pure-element opacities of nuclei with similar atomic properties \citep{2020MNRAS.493.4143F}. Specifics of the representative elements for our composition are given in Ref.~\cite{kilonova-lanl-WollaegerNewGrid2020}.

The \texttt{SuperNu} outputs are observing-angle-dependent, simulated spectra $F_{\lambda, \rm sim}$, post-processed to a source distance of $10$~pc, in units of erg s$^{-1}$ cm$^{-2}$ \AA$^{-1}$. The spectra are binned into 1024 equally log-spaced wavelength bins spanning $0.1 \leq \lambda \leq 12.8$~microns. For the purposes of this work, we consider the light curves for the Rubin Observatory $grizy$ and 2MASS $JHK$ broadband filters. As we only consider anisotropic simulations in this study, unless otherwise noted, we extract simulated light curves using 54 angular bins, uniformly spaced in $\cos\theta$ over the range $-1 \leq \cos \theta \leq 1$. The angle $\theta$ is taken between the line of sight and the symmetry axis \edited{($z=0$)} as defined in Equation~(2) of Ref.~\cite{Peng24}, with further description of the treatment of observing angle found therein. 

We interpolate over the library of \texttt{SuperNu} simulations described in \cite{Ristic22, Peng24} using the multi-layer perceptron (MLP) described in \cite{Peng24}. In summary, the MLP was trained on $\sim 250$ light-curve simulations evaluated at $264$ log-spaced times between $0.125$ and $37.24$ days for $54$ viewing angles equally spaced in $\cos\theta$ for $\theta$ ranging from $0$ to $180^\circ$. We do not perform any normalization of our inputs or outputs, with ejecta parameters ranging from $-3 \leq \log m / M_\odot \leq -1$ and $0.05 \leq v/c \leq 0.3$ and light curves ranging from -18 to 8 AB magnitudes. We train a separate MLP for each of our broadband filters, training each MLP for 1000 epochs with a batch size of 32 using the Adam optimizer. Our initial learning rate is $2~\times~10^{-4}$ with a decay rate of 5\% every 10 epochs.

\subsection{Ejecta fits to numerical relativity simulations}
\label{sec:kn_ejecta_fits}

We employ the forward model ejecta fits from Refs.~\cite{2020PhRvD.101j3002K, 2020Sci...370.1450D, 2022CQGra..39a5008N}, hereafter referred to as \KruFo, \DiCo, and \Nedora, respectively. These fits to numerical relativity simulations estimate the ejecta properties, namely dynamical and wind ejecta mass and velocity, given binary properties (mass ratio $q$ and neutron star radius $R_{\rm{1.4}}$) as inputs. In summary, \KruFo\ present fitting formulae based on 52 numerical relativity simulations of binary neutron star mergers compiled from Refs.~\cite{2018ApJ...869..130R} and \cite{2019ApJ...876L..31K}. \DiCo\ base their formulae on 73 numerical relativity simulations, using the two aforementioned sources in addition to simulations from Refs.~\cite{2017PhRvD..95d4045D} and \cite{2011PhRvD..83l4008H}. Finally, \Nedora\ employs the largest numerical relativity dataset to date for their fits, using 324 models from Refs.~\cite{2013ApJ...773...78B, 2021ApJ...906...98N, 2020PhRvD.101d4053V, 2016PhRvD..93l4046S, 2013PhRvD..87b4001H, 2018ApJ...869..130R, 2016CQGra..33r4002L, 2019ApJ...876L..31K, 2019EPJA...55..124P, 2017PhRvD..95d4045D, 2019ApJ...886L..30N, 2020MNRAS.497.1488B, 2015PhRvD..91f4059S}. The implementation of these ejecta fits follows the approach outlined in Section~II.F of Ref.~\cite{2023PhRvR...5a3168K}, with the exception of the modifications described in this section.
\edited{These fits neglect the effects of NS spin on the amount of ejected material, which can impact the amount of
  ejecta by factors of order unity for
  rapidly spinning NS \editedtwo{$(\chi \gtrsim 0.1)$}; see, e.g., \cite{2017PhRvD..95d4045D,2022PhRvD.105f4050D,2020PhRvD.102b4087C}. }

\begin{figure}[htbp!]
    \centering
    \includegraphics[width=\linewidth]{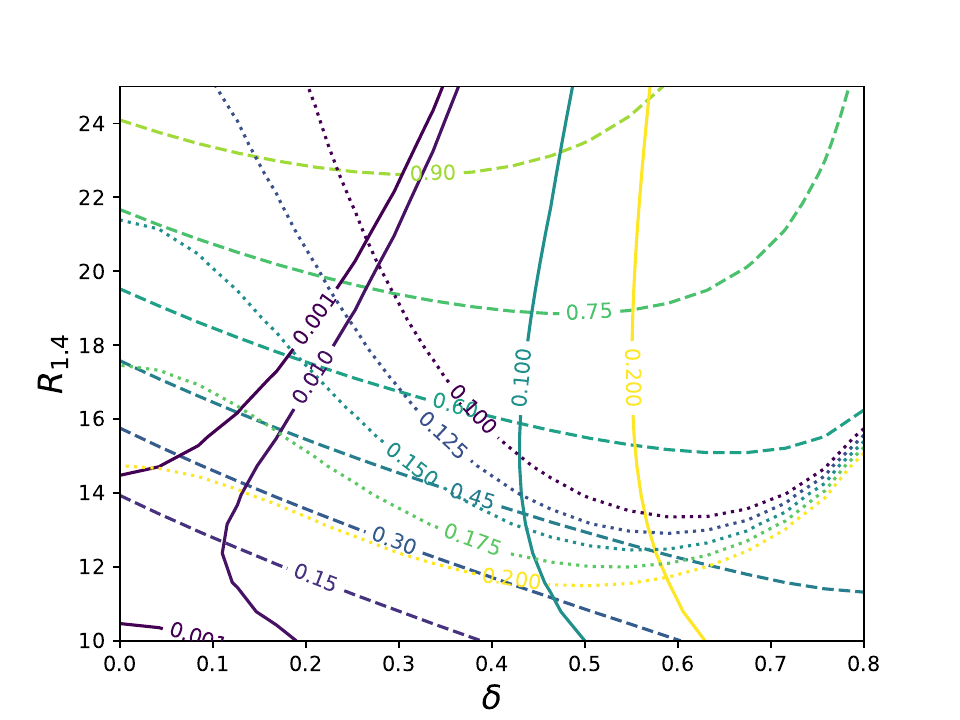}
    \caption{Solid, dashed, and dotted line contours corresponding to the \KruFo\ dynamical ejecta mass $m_{\rm{ej}}$,
      disk mass $m_{\rm{disk}}$, and ejecta velocity $v_{\rm{ej}}$, respectively. Contours were calculated as a function
      of mass asymmetry $\delta = (m_1  -m_2)/(m_1+m_2)$ and neutron star radius $R_{\rm{1.4}}$. Mass and velocity units are in $M_\odot$ and $c$, respectively.}
    \label{fig:contours}
\end{figure}

Each ejecta fit predicts the ejecta mass, \editedtwo{total} ejecta velocity, and accretion disk mass as a function of the binary mass ratio $q$ and the neutron star radius $R_{\rm{1.4}}$.
\edited{We relate $R_{\rm{1.4}}$ to $\tilde{\Lambda}$ as $R_{\rm{1.4}} \simeq 13.4 (\tilde{\Lambda}/800)^{1/6}$ km per Eq. (18) in Ref. \cite{2018PhRvD..98f3020Z}}. In none of the ejecta fit models is there a prescription for tying the disk mass to post-merger wind mass; instead, we map the fits' ejecta mass prediction to the dynamical ejecta, and assume some fraction \fdisk of the disk mass is ejected as wind ejecta.
In Figure~\ref{fig:contours}, we display naive contours corresponding to representative values of these ejecta
parameters as a function of the binary parameters, fixing the binary source-frame chirp mass to the observed value for GW170817.
The solid, dashed, and dotted lines indicate $m_{\rm{ej}}(\delta, R_{\rm{1.4}})$, $m_{\rm{disk}}(\delta, R_{\rm{1.4}})$, and $v_{\rm{ej}}(\delta, R_{\rm{1.4}})$, respectively, \edited{where $\delta = (m_1  -m_2)/(m_1+m_2) = (1-q)/(1+q)$ indicates the mass asymmetry in the binary}.
\editedtwo{For reference, for GW170817-like systems as the one shown in Figure \ref{fig:contours}, $\delta \sim 0.1 \implies q \sim 0.81$ while $\delta \sim 0.2 \implies q \sim 0.66$. The $\delta$ space explored in our study allows for a substantial fraction of extremely asymmetric configurations, which we will show are necessarily explored by the ejecta fits to reconcile EM and GW observations (assuming the constraints presented by our kilonova model) without the use of additional astrophysical priors.}
The solid dynamical ejecta contours constrain the mass ratio $q$ while allowing for a broad range of radii. 
Conversely, the dashed disk mass contours constrain $R_{\rm{1.4}}$ to favor low values for realistic disk masses \edited{($m_{\rm{disk}}/M_\odot < 0.45$, as in, e.g., Ref. \cite{2023ApJ...945L..13C})} with a slight constraint on $\delta$.
Critically, however, the dotted $v_{\rm{ej}}$ contours -- shown at levels corresponding to our previously-inferred
estimate ($0.10-0.20c$, \cite{Ristic22, 2023PhRvR...5d3106R}) for this quantity -- exclude the conventional interpretation for GW170817, which consists of  $\delta \simeq 0$ (comparable masses) and
$R_{1.4} \simeq 12\unit{km}$.  Instead, the $v_{\rm {ej}}$ contours follow a trendline that permits either comparable
mass but large radius or conventional radius but highly asymmetric masses, both scenarios that could produce
sufficiently low  dynamical ejecta velocities in binary neutron star mergers.

Because electromagnetic observations constrain these three ejecta properties independently (by way of kilonova models which introduce their own systematic uncertainties), only in exceptional
circumstances can all three equations be solved simultaneously for a self-consistent pair of binary properties $(\delta,
R_{1.4}$: we have more equations than unknowns).  However, the ejecta parameters preferred by our previous analyses,
notably and critically the ejecta velocity, are at or beyond the range of underlying simulation data used to train these
analytic approximate fits.  As a result, motivated by previous work, we generously allow for independent systematic uncertainty in
all three predictions for ejecta properties.

Rather than estimating the uncertainties on the ejected mass from each of our two components, we assign free parameters $\alpha_{\rm{dyn}}$ and $f_{\rm{disk}}$ to quantify these systematic uncertainties during inference. 
The $\alpha_{\rm{dyn}}$ parameter serves as a scale factor for the fits' dynamical ejecta mass prediction, while $f_{\rm{disk}}$ dictates what fraction of the accretion disk is expelled as wind ejecta.
Given the behavior of the dotted $v_{\rm{ej}}$ contours in Figure~\ref{fig:contours}, we also \edited{introduce} a systematic uncertainty parameter for the ejecta velocity, $\beta_{\rm\phi}$. 
To uniformly explore the systematic uncertainty in velocity space, we apply the reparameterization $\phi = \ln((c/v)^2 -1)$.
Our uncertainty is  characterized by a uniformly distributed random variable $\beta_\phi$ added to $\phi$.

The ejecta fits predict a total ejecta velocity $v_{ej}$ by adding in quadrature the velocities describing the ejecta within and perpendicular to the orbital plane, \edited{without a direct one-to-one mapping to the standard kilonova dynamical and wind ejecta components} (see Eqs.~(5)--(9) in Ref.~\cite{2017CQGra..34j5014D}). In this study, \edited{to ensure maximal consistency with established methodology in the literature, we inherit} previous conventions. As in Ref. \cite{2018MNRAS.480.3871C}, we set the dynamical ejecta velocity to \editedtwo{that of the total ejecta velocity such that $v_d = v_{ej}$} and fix the wind velocity $v_w = 0.10c$ per the results of our previous analysis in Ref. \cite{Peng24}. The choice to fix $v_w$ is similar to previous approaches, e.g. as in Ref.~\cite{2023PhRvR...5a3168K}, where the wind velocity $v_w$ was set to $0.08c$ as informed by Ref.~\cite{2018ApJ...869..130R}. 
\editedtwo{We caution readers that in attributing $v_{ej} = v_d$ as a makeshift connection between binary and ejecta parameters, we stray away from the expected evolution of key quantities (e.g. matter density, temperature) as a function of their total expansion velocity.}

The employed fits also have a lower limit on the accretion disk mass of $\log M_{\rm{disk}} \ge -3$, but no defined upper limit. As certain combinations explored during sampling may yield \edited{ejecta} parameters outside of the realm of validity \edited{of our surrogate model}, we set an associated restriction on the wind mass such that $\log m_w \le -1$. \editedtwo{The same upper limit applies to the dynamical ejecta to ensure ejecta samples stay within the region of validity.}

\subsection{The likelihood of ejecta parameters for AT2017gfo given electromagnetic observations}
\label{sec:em_likelihood}

The AT2017gfo data is originally presented in \cite{2017PASA...34...69A, 2017Natur.551...64A, 2017Sci...358.1556C,
  2017ApJ...848L..17C, Diaz17, 2017Sci...358.1570D, Evans_2017, 2017SciBu..62.1433H, 2017Sci...358.1559K,
  2017ApJ...850L...1L, 2017Natur.551...67P, 2018ApJ...852L..30P, shappee2017early, 2017Natur.551...75S,
  2017ApJ...848L..27T, 2017Natur.551...71T, 2017PASJ...69..101U}.
Each \edited{combination of ejecta parameters} $\vec{x}_{\theta}$ is evaluated by the {MLP} to produce a light-curve prediction $\hat{y}$ for every one of the \emph{grizyJHK} broadband filters. We calculate the residual between the MLP prediction $\hat{y}$ and the AT2017gfo observed data in time $i$ for every band $B$ by way of the reduced-$\chi^2$ statistic

\begin{equation}
    \chi^2 = \sum_{i, B} \frac{(\hat{y}_{i, B} - d_{i, B})^2}{\sigma_i^2 + \sigma_{\rm sys}^2}.
\end{equation}

In our $\chi^2$ residual calculation, we include statistical uncertainties from the AT2017gfo data $\sigma_i$, as well
as systematic uncertainties $\sigma_{\rm{sys}}$ which we use as a catch-all term to encompass all uncertainties,
quantifiable or otherwise, associated with the neural network interpolation process and kilonova modeling uncertainties.  Unlike in previous work (Ref.~\cite{Peng24}), we set our systematic modeling uncertainty
$\sigma_{\rm{sys}}$ as a free parameter to allow for easier reconciliation of the GW and EM inferences. We adopt a purely Gaussian likelihood based on these residuals, i.e.
\begin{equation}
\ln {\cal L_{\rm{EM}}} = -\frac{\chi^2}{2} - \frac{1}{2}\ln (2\pi)^N\sum_i(\sigma_i^2+\sigma_{\rm sys}^2) \,,
\end{equation}
where $N$ is the total number of AT2017gfo observations. 

\subsection{The likelihood of source parameters for GW710817 given gravitational wave observations}
\label{sec:gw_likelihood}

Following previous studies that constrain properties of GW170817 and the equation of state from GW observations \cite{LIGO-GW170817-SourceProperties,LIGO-GW170817-EOSrank,gwastro-bns-eos-Atul2024,gwastro-ns-eos-Vilkha2024}, we use the RIFT parameter
inference engine \cite{gwastro-PENR-RIFT,gwastro-PENR-RIFT-GPU,gwastro-RIFT-Update}  to explore and
evaluate the marginal likelihood $L(x)$ for a wide range of different compact binary source parameters $x$.  Following the RIFT
paradigm, the parameters $x$ include all intrinsic (detector-frame) quantities needed to phenomenologically characterize
the binary's inspiral: its two component masses, spins, and tidal deformabilities, assumed independent of any specific
EOS model.
We then use standard nonparametric interpolation techniques (here, random forests from
\texttt{sklearn}) to interpolate the marginal likelihood to provide a continuous estimate $\hat{L}(x)$. \edited{Our GW likelihood takes the form
\begin{equation}
    \ln \mathcal{L}_{\rm{GW}} = \ln \hat{L} ( \mathcal{M}_c, q, \chi_{\rm{eff}}, \tilde{\Lambda}),
\end{equation}
with input parameters derived from $x~=~(m_1, m_2, \chi_1, \chi_2, \Lambda_1, \Lambda_2)$.}
To be more concrete, we will employ precisely the same analysis framework and settings as used in Ref. \cite{gwastro-ns-eos-Kedia2024}.  As a brief review to establish notation, gravitational wave observations of GW170817 constrain the binary masses $m_i$ and tidal deformabilities $\Lambda_i$ for
each component $i$ of the binary
\cite{LIGO-GW170817-bns,LIGO-GW170817-SourceProperties,LIGO-GW170817-EOS,PhysRevResearch.2.043039}. The presence of matter  impacts the binary's inspiral at leading order through each component's dimensionless tidal deformability parameter  $\Lambda_i = (2/3) k_2 (c^2 r_i / G m_i)^5$~\cite{Flanagan:2007ix}, where $k_2$ is the $l = 2$ Love number, $m_i$ is the neutron star mass, and $r_i$ is the radius of a $1.4 M_\odot$ neutron star.
The leading order contribution to the phase evolution of a GW inspiral is given by the weighted combination of
$\Lambda_i$ terms
\begin{align}
\label{eq:lt}
    \tilde{\Lambda} = \frac{16}{13} \frac{(m_1 + 12 m_2) m_1^4 \Lambda_1 + (m_2 + 12 m_1) m_2^4 \Lambda_2}{(m_1 + m_2)^5} \,.
\end{align}
The impacts of tidal effects on outgoing radiation are included in many conventional approximate phenomenological
estimates of outgoing radiation from merging compact binaries.   In this work, we begin with precisely the marginal
likelihood data accumulated in Ref.~\cite{gwastro-ns-eos-Kedia2024}. That investigation was performed with a
contemporary state-of-the-art model, IMRPhenomPv2\_NRTidalv2 \cite{2019PhRvD..99b4029D,2019PhRvD.100d4003D}, which
incorporates precession physics but omits higher-order modes.  
This \editedtwo{GW} analysis was performed 
with open GW data for GW170817 available  from GWOSC \cite{ligo-O1O2-opendata%
}, using the same power spectral densities provided with GWTC-1  \cite{LIGO-O2-Catalog,LIGO-O2-Catalog-PSDRelease}, over a
frequency range from $23\unit{Hz}$ to $1700\unit{Hz}$, %
with known sky location and source
luminosity distance derived from the electromagnetically-identified host galaxy.
For exploration purposes, most prior assumptions were customary
(e.g., uniform in detector-frame component masses), noting that the previous investigation employed multiple
spin priors and mass ratio priors to accumulate marginal likelihood information over a wide range of source parameters.
Although this analysis included prior information about the alignment between the binary's angular momentum direction and the line of sight,
inferred from late-time radio afterglow observation \cite{2019NatAs...3..940H, 2018Natur.554..207M}, this information
should not significantly impact \editedtwo{the GW} intrinsic parameters \editedtwo{(the binary component masses, spins, and tidal deformabilities)}. 

Because the original marginal likelihood grid accumulated in Ref.~\cite{gwastro-ns-eos-Kedia2024} did not extend to
cover all of the extreme BNS configurations needed for this study, we replicated and extended this analysis to ensure that
marginal likelihood data was present whenever required for joint EM/GW analysis.
For simplicity, we employ the same GW model throughout for all $\tilde{\Lambda}$, keeping in mind that this model was not
calibrated over the full range of $\Lambda$ needed to carry out our investigation.

\begin{figure}[htbp!]
    \centering
    \includegraphics[width=\linewidth]{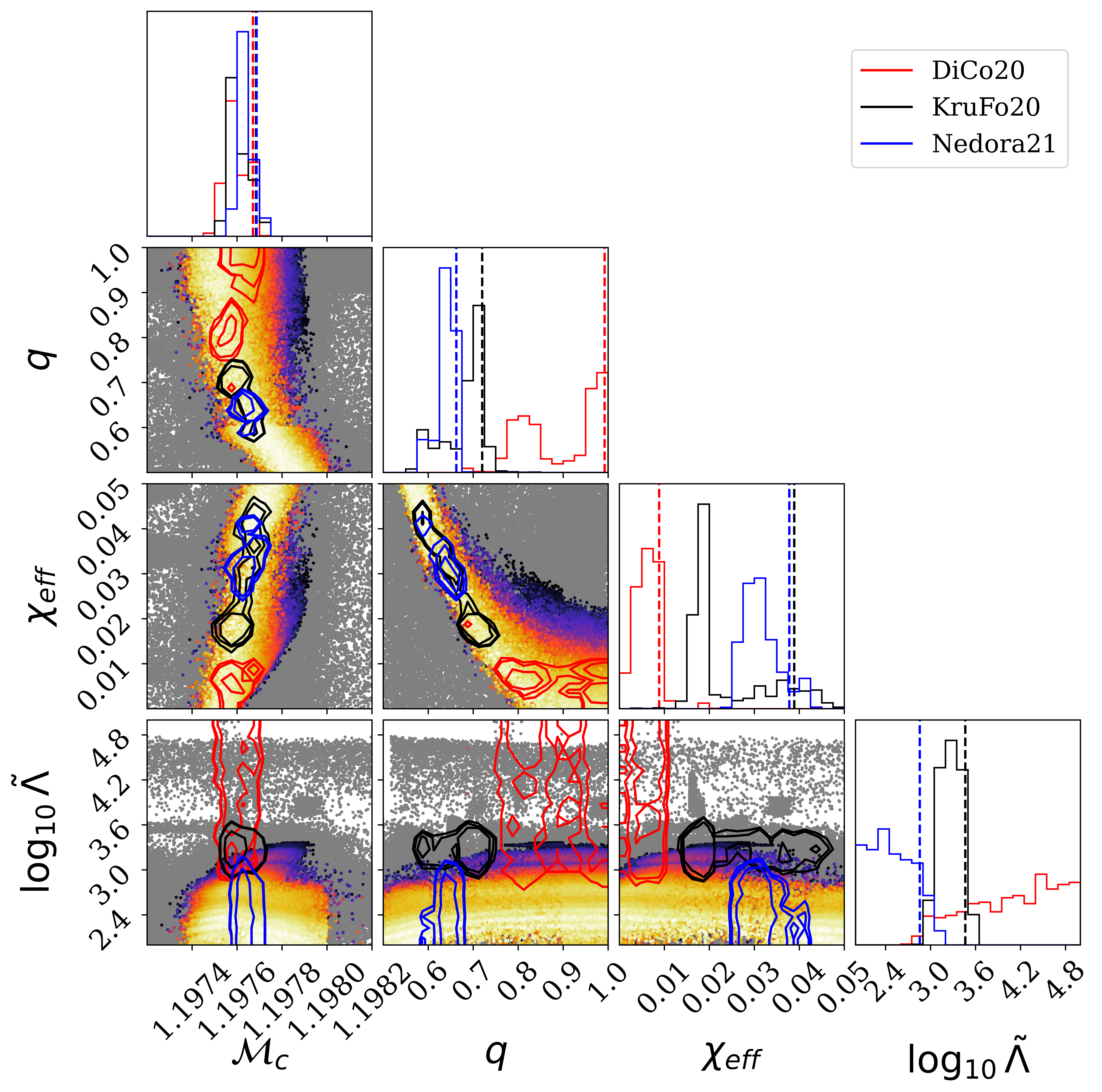}
    \caption{In this figure, the colored dots represent the input marginal likelihood data assumed in our analysis. In black, red, and blue, we overplot posteriors corresponding to the $90\%$ credible interval for the \KruFo, \DiCo, and \Nedora\ GW+EM predictions, respectively. Here, $\mathcal{M}_{\rm{c}}$ is in the detector frame.}
    \label{fig:GW_parameter_posterior}
\end{figure}

Figure~\ref{fig:GW_parameter_posterior} shows the raw marginal GW likelihood data employed in our analysis, before
continuous interpolation over binary neutron star parameters. We show the GW likelihood as a function of the chirp mass $\mathcal{M}_c = (m_1m_2)^{3/5}/(m_1+m_2)^{1/5}$, mass ratio $q = m_2/m_1$, effective spin $\chi_{\rm{eff}} = (a_{\rm{1z}} + qa_{\rm{2z}})/(1+q)$, and effective tidal deformability $\tilde{\Lambda}$ as defined in Equation~\ref{eq:lt}. In this figure,
the color indicates the marginal likelihood, scaling from higher (white) to lower (black) likelihood values, with gray
points showing values below the $\ln \mathcal{L}_{\rm{max}} - 15$ cutoff.
This figure highlights the broad coverage of our underlying likelihood data, particularly the wide range of
$\tilde{\Lambda}$.  Our coverage includes carefully assessing the marginal likelihood of GW signals in many regions far
outside the customary GW posterior, which is concentrated in the white region.   To put our broad exploration in
context, drawing upon analyses to be discussed in detail in Section~\ref{sec:all_systematics}, in this figure we also
show the results of three joint GW+EM inferences, using the \DiCo\ (red), \KruFo\ (black), and \Nedora\ (blue)  ejecta
fits, respectively.  As described later, all three analyses adopt  conventional spin
and mass ratio priors, consistent with previously published
interpretations using comparable models, settings, and data \cite{LIGO-GW170817-bns,LIGO-GW170817-SourceProperties}.
Nonetheless, our  joint GW+EM inferences favor more extreme binary and/or tidal parameters compared to previously
reported posterior inferences using GW alone, including asymmetric mass ratio (all except \DiCo),  large tidal
deformability (\KruFo\ and \DiCo), and sometimes even large spin (\Nedora).
As these extreme configurations are not well-represented in widely-used fiducial posterior samples from the original
analyses, previous follow-up GW+EM investigations that rely on simply reweighting these fiducial samples will not fully
capture the features that we report below.
In other words, our subsequent calculations demand the wide-ranging likelihood-based approach adopted here.

\subsection{Bayesian inference}
\label{sec:inference}

We use the RIFT framework \cite{Wofford2023} to adaptively perform Monte Carlo integration and generate samples \edited{using a joint likelihood $\ln \mathcal{L} = \ln \mathcal{L}_{\rm{GW}} + \ln \mathcal{L}_{\rm{EM}}$}. In summing the separate GW and EM likelihoods, we assume their equal contribution to the joint likelihood and that our error model is well understood and lacking additional systematics (both modeling and data). While possible choices for composite likelihood weights have been discussed in the literature \cite{2024arXiv240402313L}, in this work we stick to established conventions and leave the exploration of weighted composite likelihoods to a separate study.
As previously, we employ an adaptive volume Monte Carlo integrator, following closely the approach outlined in Ref.~\cite{2023PhRvD.108b3001T}. 
The adaptive volume integrator allows for more efficient sampling given the higher-dimensional space explored in this work. 
As our inference is performed via adaptive Monte Carlo integration, the reliability of our posterior can be expressed in terms of a number of effective samples $n_{\rm eff}$. 
Several different conventions exist for this number; see the appendix of Ref.~\cite{gwastro-RIFT-Update} for more detailed discussion. 
For this study, we terminate our analyses when $n_{\rm eff} \simeq 10^3$, indicating sufficient convergence.

Table~\ref{tbl:priors} shows the sampling parameters used during inference, along with their associated lower and upper limits and priors. 
The combination of the chirp mass ${\cal M}_c $ and mass asymmetry $\delta$ priors provide an equivalent locally uniform prior for the binary masses $m_1$ and $m_2$. 
In addition to the masses, we also sample for the spins of the neutron stars $s_{\rm{1z}}$ and $s_{\rm{2z}}$. 
We assume that both of the neutron stars have the same radius given a 1.4~$M_\odot$ neutron star $R_{1.4}$, treating this radius as a proxy for the nuclear EOS. 
As described in Sec.~\ref{sec:kn_ejecta_fits}, we introduce parameters $\alpha_{\rm{dyn}}$, $f_{\rm{disk}}$, and $\beta_\phi$ that encapsulate the uncertainty in our dynamical ejecta mass, wind ejecta mass, and dynamical ejecta velocity, respectively. 
These parameters are sampled in addition to the light-curve systematics $\sigma_{\rm{sys}}$, effectively allowing for three separate measures of systematic uncertainty in our analysis. 
Finally, we sample for the viewing angle $\theta$ as in prior work (e.g. Ref.~\citep{Peng24}). 
All parameters in Table~\ref{tbl:priors} are assumed to be dimensionless unless their names are followed by units in brackets.
\begin{table}[htbp!]
    \centering
    \begin{tabular}{c|c|c}
         Parameter & Limits & Prior \\ \hline
         $\mathcal{M}_{c}$ {[$M_\odot$]} & {[1.1853, 1.1880]} & Jointly uniform $m_1,m_2$ \\
         $\delta$ & ${[0, 0.8]}$ & Jointly uniform $m_1,m_2$\\
         $s_{\rm{1z}}$ & {[$0, 0.05$]} & Uniform \\
         $s_{\rm{2z}}$ & {[$0, 0.05$]} & Uniform \\
         $R_{1.4}$ {[km]} & {[$8, 50$]} & Uniform \\
         $\log \alpha_{\rm{dyn}}$ & {[$-5, 5$]} & Uniform \\
         $\log f_{\rm{disk}}$ & {[$-5, 0$]} & Uniform \\
         $\beta_\phi$ & {[$-2, 3$]} & Uniform \\
         $\sigma_{\rm{sys}}$ & {[0, 8]} & Uniform \\
         $\theta$ {[deg]} & {[$0, 90$]} & $\mathcal{N}(\mu=20, \sigma=5)$
    \end{tabular}
    \caption{Sampling parameters along with their associated limits and priors. Parameters are dimensionless unless
      units in brackets are specified in the parameter column.
The priors in ${\cal M}_c $ and $\delta$ are equivalent to a locally uniform probability density in $m_1,m_2$. }
    \label{tbl:priors}
\end{table}

\section{Results}
\label{sec:results}

\editedtwo{For all the analyses in this section, we assume that the dynamical ejecta mass systematic uncertainty $\alpha_{\rm{dyn}}$ and light-curve systematic uncertainty \sigmasys are treated as free parameters.
In Section \ref{sec:all_systematics}, we treat $f_{\rm{disk}}$ and $\beta_\phi$ as free parameters to characterize the uncertainties associated with the wind ejecta mass (derived from the ejected disk mass) and ejecta velocity, respectively.
In Section \ref{sec:fixed_beta_phi}, we fix $\beta_\phi = 0$ to show the effect of ignoring ejecta velocity systematics during inference.
Finally, in Section \ref{sec:fixed_f_disk}, we fix both $\beta_\phi = 0$ and $f_{\rm{disk}} = 0.30$ to highlight how making assumptions about no ejecta velocity or disk ejection fraction systematics renders meaningful interpretation of our inference results intractable.}

\subsection{All systematics included \texorpdfstring{($\alpha_{\rm{dyn}}, f_{\rm{disk}}, \beta_\phi$ free)}{adyn, fdisk, bf}}
\label{sec:all_systematics}

\begin{figure}[htbp!]
    \centering
    \includegraphics[width=\linewidth]{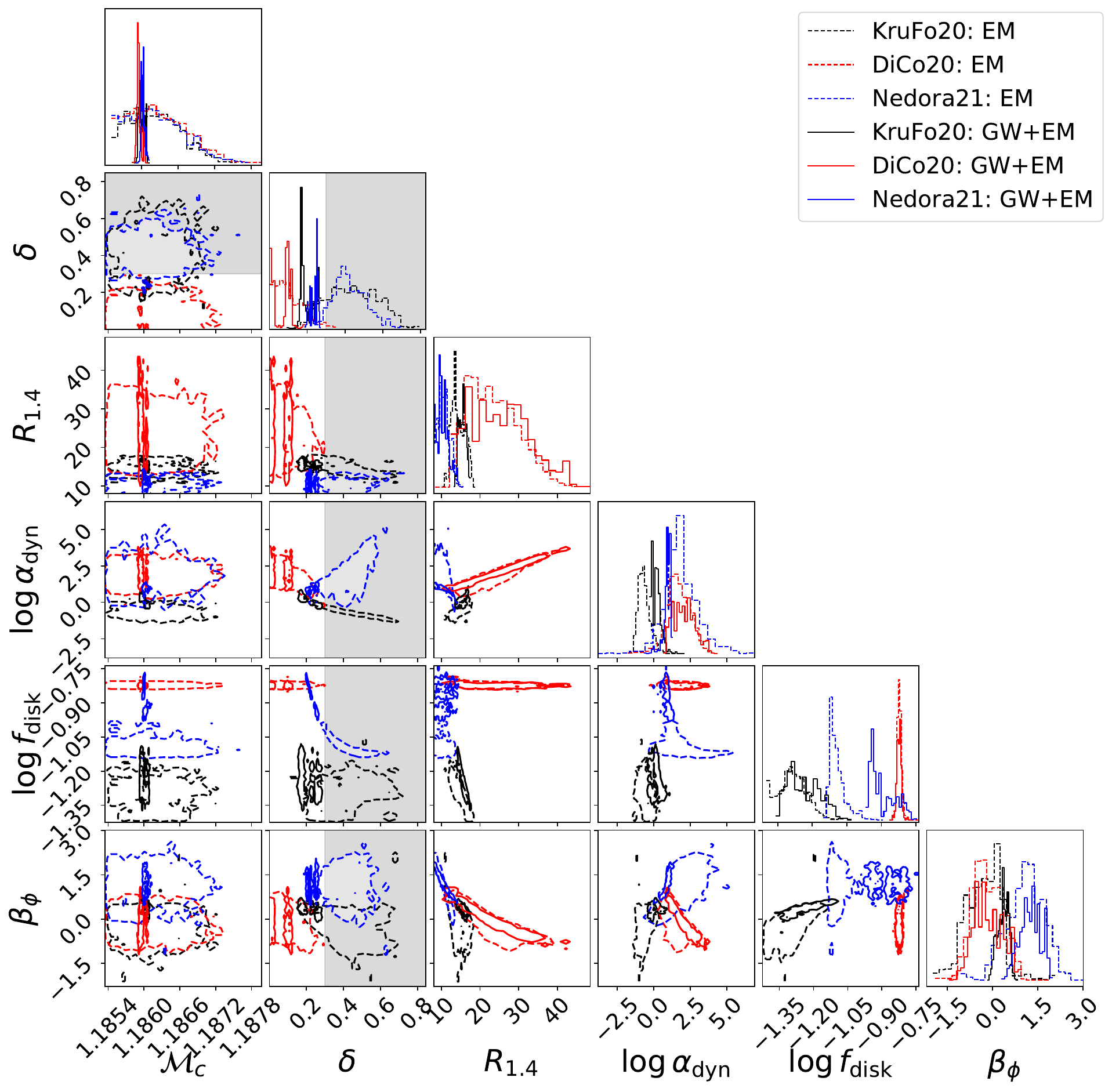}
    \caption{Binary parameter posterior distributions using EM (dashed) and GW+EM (solid) likelihoods using $\log f_{\rm{disk}}$ priors of [-0.9, -0.75], [-1.4, -1.2], and [-1.2, -1] for \DiCo, \KruFo, and \Nedora, respectively. The inclusion of the GW likelihood most noticeably affects the chirp mass $\mathcal{M}_c$ and mass asymmetry $\delta$, while simultaneously providing tighter constraints on the neutron star radius $R_{\rm{1.4}}$ and ejected disk fraction $f_{\rm{disk}}$. \editedtwo{Gray shaded regions represent $\delta > 0.3$ for which the secondary star's mass is sub-solar ($< 1 M_\odot$).}}
    \label{fig:corner_binary_narrow_mc_fix_vw}
\end{figure}

The corner plots in Figure~\ref{fig:corner_binary_narrow_mc_fix_vw} show 90\%
two-dimensional credible intervals and one-dimensional marginal posterior distributions for the underlying binary
($\mathcal{M}_c,\delta$), EOS ($R_{\rm{1.4}}$), and ejecta systematic ($f_{disk},\alpha_{dyn}, \beta_\phi$) model parameters. 
The dashed black, red, and blue lines correspond to inferences using only AT2017gfo data $\mathcal{L}_{\rm{EM}}$ for the \KruFo, \DiCo, and \Nedora\ ejecta fits, respectively. 
The solid lines show the same analysis when incorporating both the AT2017gfo $\mathcal{L}_{\rm{EM}}$ and GW170817 $\mathcal{L}_{\rm{GW}}$ data.
We do not plot the posteriors for $s_{\rm{1z}}, s_{\rm{2z}}$, and $\sigma_{\rm{sys}}$ as they are largely uninteresting in the context of understanding the ejecta fits and would have reduced the legibility of the figure.
Following a preliminary analysis to identify the posterior support for each ejecta fit,  we perform focused analyses
targeting narrow regions of  $f_{\rm{disk}}$ for each ejecta fit:  $\log f_{\rm{disk}}$ is uniform over [-0.9, -0.75], [-1.4, -1.2], and [-1.2, -1] for \DiCo, \KruFo, and \Nedora, respectively.

\begin{figure}[htbp!]
    \centering
    \includegraphics[width=\linewidth]{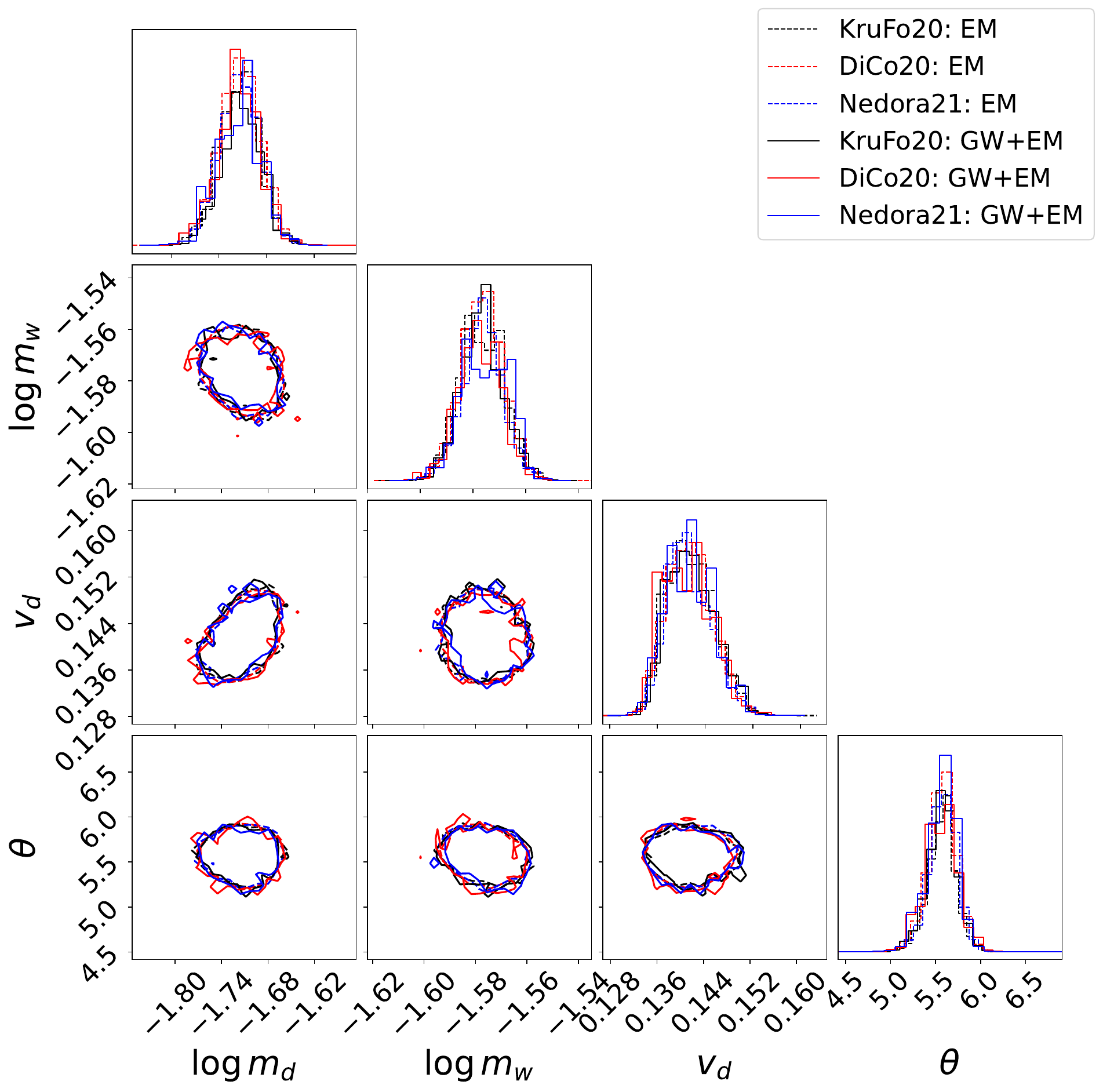}
    \caption{Ejecta posterior distributions using EM (dashed) and GW+EM (solid) likelihoods with the disk mass ejection fraction left as a free parameter. The wind velocity is fixed to $v_w = 0.10c$ per the posteriors in Ref. \cite{Peng24}. There is little difference in the ejecta posteriors between the two approaches, modulo artifacts from varying sample counts, due to our kilonova model requiring parameters in a very narrow region to recreate the AT2017gfo light curves. Figure \ref{fig:corner_binary_narrow_mc_fix_vw} highlights the differences in the binary parameters when including the GW likelihood.}
    \label{fig:corner_ejecta_narrow_mc_fix_vw}
\end{figure}

We verify that our ejecta fit posteriors produce consistent kilonova ejecta properties in Figure~\ref{fig:corner_ejecta_narrow_mc_fix_vw}. 
We begin by examining the posteriors for the binary's ejecta properties, primarily constrained by our EM likelihood $\mathcal{L}_{\rm{EM}}$. 
Figure~\ref{fig:corner_ejecta_narrow_mc_fix_vw} shows the predicted ejected masses, bulk velocity (attributed to the dynamical ejecta), and the viewing angle derived from the inferences presented in Figure~\ref{fig:corner_binary_narrow_mc_fix_vw}.
In this analysis, we adopt $v_w=0.10c$, \edited{consistent with the viscosity-driven wind ejecta} in numerical simulations and our previous analysis of AT2017gfo \cite{Peng24}.
As in Figure \ref{fig:corner_binary_narrow_mc_fix_vw}, the dashed curves represent inference using data from AT2017gfo alone and the solid curves consider both the AT2017gfo and GW170817 data. 
Due to considerable flexibility in ejecta mass model systematics (i.e., $f_{\rm
  {disk}},\alpha_{\rm{dyn}}, \beta_\phi$), our model recovers once again a nearly indistinguishable distribution of ejecta parameters. 
Since our inference tries to fit AT2017gfo observations with our model, and since our ejecta model allows for systematic error, 
our inferences always find consistent solutions for the best-fitting ejecta, i.e. 
the ejecta parameters that consistently explain AT2017gfo within the context of our kilonova model family, subject to the limitations of those kilonova models. 
The much larger and more rapidly varying electromagnetic likelihood dominates our inference of these ejecta properties. By
contrast, the GW information only slightly perturbs the overall conclusions obtained electromagnetically.
The inclusion of an ejecta velocity systematic parameter does not affect the similarity of the ejecta posteriors.
\editedtwo{As in Ref. \cite{Peng24}, our $\theta$ posteriors deviate appreciably from the prior, owing to our model reconciling the tension between the ``blue" and ``red" emission by sampling away from the $\theta$ prior (see their Fig. 10).}

\begin{figure}[htbp!]
    \centering
    \includegraphics[width=\linewidth]{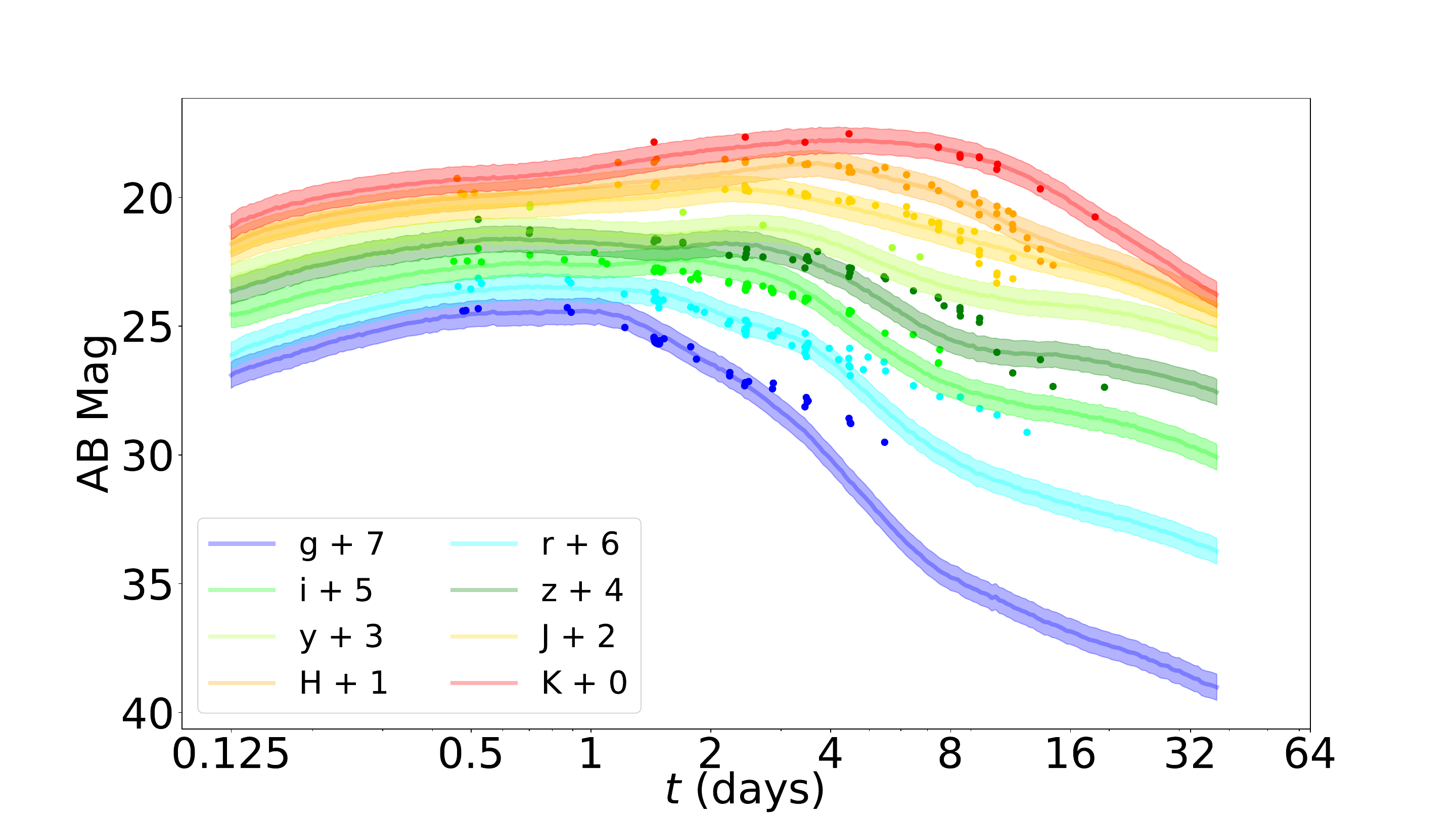}
    \caption{Light curves predicted by our kilonova surrogate model for the \Nedora\ GW+EM ejecta posteriors presented
      in Figure~\ref{fig:corner_ejecta_narrow_mc_fix_vw}, with solid lines corresponding to the median posterior values
      and shaded bands representing 1$\sigma$ uncertainty. This posterior includes the effect of
        marginalizing over unknown model systematic uncertainty $\sigma_{\rm sys}$, as described in the text.}
    \label{fig:lc_unconstrained}
\end{figure}

In Figure~\ref{fig:lc_unconstrained}, we plot the light curves associated with the \Nedora\ GW+EM posteriors from
Figure~\ref{fig:corner_ejecta_narrow_mc_fix_vw}, showing both the mean and posterior 68\% credible interval.   The width
of these posterior intervals derives from and reflects the narrow posteriors shown in
Figure~\ref{fig:corner_ejecta_narrow_mc_fix_vw} (including suppressed parameters reflecting model systematic uncertainty). 
As in previous studies (e.g. \cite{Ristic22, 2021ApJ...913..100K}), we observe residual model systematics relative
to the data,  where the largest deviations between our models and the data occur in the bluer filters, namely the $g$ and $r$ bands, starting at $\sim 3$ days.
Though we explore possible reasons for this blue filter discrepancy in prior work (see Figure 2 of \cite{Ristic22}), we do not explicitly include this model fidelity issue in our formulation of $\sigma_{\rm{sys}}$.
\edited{We note that a better fit to late-time blue photometry would most likely invoke a slightly more massive, slower wind ejecta component; a slower wind would produce a longer-lived blue kilonova, with the increased mass accounting for the flux loss from slower-moving ejecta (assuming ejecta under local thermodynamic equilibrium (LTE), see Ref. \cite{2025arXiv250818364B} for non-LTE effects on late-time emission).}
Both in this work, where we treat \sigmasys as a free parameter, and in our previous EM-only inferences, where we assume $\sigma_{\rm{sys}}$~=~0.5 \cite{Ristic22, gwastro-nsnuc-kilonova-Ristic-InterpSpectra2023, Peng24}, the deviations that would arise from variations in the posteriors are well-encompassed by our systematics.

As expected, our analysis finds that the three ejecta models can only explain AT2017gfo using dramatically different progenitor binaries and EOS. 
Two of the three models prefer that the progenitor of AT2017gfo is a highly asymmetric binary ($\delta > 0.2 \implies q < 0.66$) with smaller NS radius $R_{\rm{1.4}}$, with the \DiCo\ models allowing equal-mass ($\delta = 0$) binaries with exceptionally large NS radius.
Conversely, these inferences suggest that a surprisingly small fraction ($\sim 4-20\%$) of
the expected disk mass is ejected as a wind, in contrast to most previous work \cite{2020ARNPS..70...95R, 2020PhRvD.101h3029F}, but consistent with predictions from three-dimensional general
relativistic magnetohydrodynamic (GRMHD) disk simulations with full transport neutrino radiation \cite{2019PhRvD.100b3008M} \edited{(although past 1 second, these simulations have a larger fraction of ejected mass \cite{2024ApJ...962...79S}; more late-time ($t > 1$ sec) disk simulations are required for a conclusive comparison)}.

\subsection{Allowing no uncertainty in ejecta velocity (\texorpdfstring{$\alpha_{\rm{dyn}}, f_{\rm{disk}}$ free, $\beta_\phi = 0$}{Bf = 0})}
\label{sec:fixed_beta_phi}

\begin{figure}[htbp!]
    \centering
    \includegraphics[width=\linewidth]{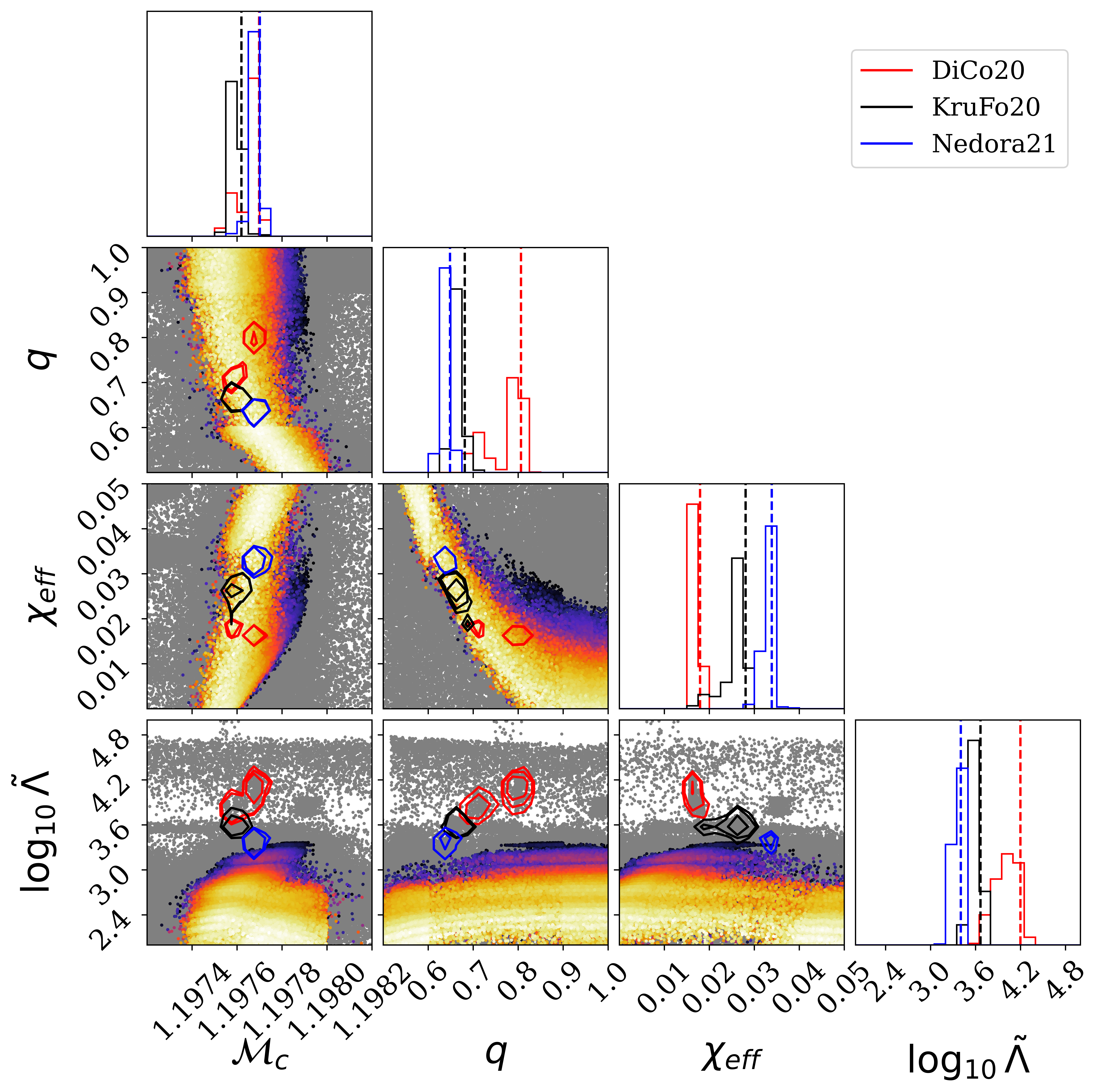}
    \includegraphics[width=\linewidth]{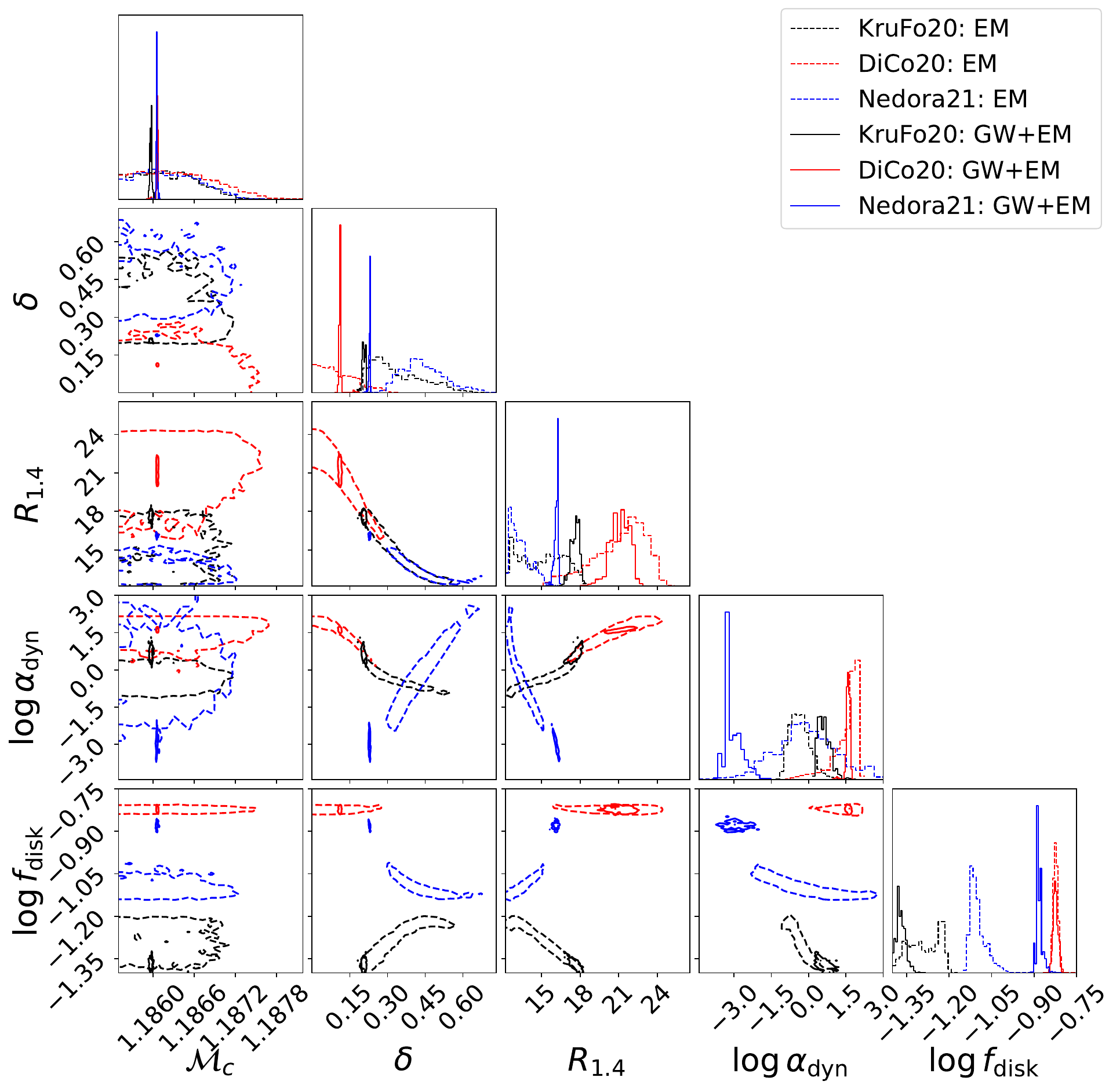}
    \caption{Same as Figures~\ref{fig:GW_parameter_posterior} and \ref{fig:corner_binary_narrow_mc_fix_vw}, but not allowing for systematic uncertainty in $v_d$ (i.e. $\beta_\phi = 0$). Note the recreation of the $v_{\rm{ej}}$ contours from Figure~\ref{fig:contours} in the $R_{\rm{1.4}}$ vs. $\delta$ plane.}
    \label{fig:corner_binary_narrow_mc_fix_vw_nobf}
\end{figure}

We perform an additional suite of studies in which we assume no systematic uncertainty in the $v_{\rm{ej}}$ predicted by the ejecta fits. 
For these analyses, we continue to treat $f_{\rm{disk}}$ as a free parameter.
Despite the assumption of no velocity systematic uncertainty made in this section, the ejecta posteriors remain unchanged, behaving as in Figure~\ref{fig:corner_ejecta_narrow_mc_fix_vw}.
Thus, we still find that our kilonova model and subsequent EM likelihood provide the dominant constraint for the ejecta parameters.

However, without an additional parameter to encapsulate the ejecta systematics, we find several notable differences,
largely dictated by the $v_{\rm{ej}}$ contours in Figure~\ref{fig:contours}.
Specifically, we find that for all three ejecta fits, the GW+EM analyses all favor large radii in order to replicate the
low dynamical ejecta velocities necessitated by the $v_{\rm{ej}} = v_d$ posteriors in
Figure~\ref{fig:corner_ejecta_narrow_mc_fix_vw}.
Based on Figure~\ref{fig:contours},  a tight constraint on $v_{\rm ej}$ requires either asymmetric NS \edited{closer to more conventional} %
radii ($\delta \simeq 0.4$ and $R_{1.4}\simeq 15\unit{km}$) or symmetric NS with extreme radii.
Indeed, the union of all three $\delta-R_{1.4}$ posteriors in Figure~\ref{fig:corner_binary_narrow_mc_fix_vw_nobf} closely
mirrors the underlying $v_{\rm ej}$ contours.  As a result, this constraint forces the posterior away from the peak GW
likelihood, which favors lower NS radii and more symmetric binaries.  In all three cases where we omit systematic error
in velocity ($\beta_\phi=0$), joint GW+EM inference mildly favors
asymmetric binaries $\delta \simeq 0.1-0.2$  with large NS radius $R_{1.4}>15\unit{km}$.

\subsection{Allowing no uncertainty in ejecta velocity and disk mass ejection fraction (\texorpdfstring{$\alpha_{\rm{dyn}}$ free, $\beta_\phi$}{Bf} = 0 and \texorpdfstring{$f_{\rm{disk}} = 0.30$}{fdisk = 0.30})}
\label{sec:fixed_f_disk}

\begin{figure}[htbp!]
    \centering
    \includegraphics[width=\linewidth]{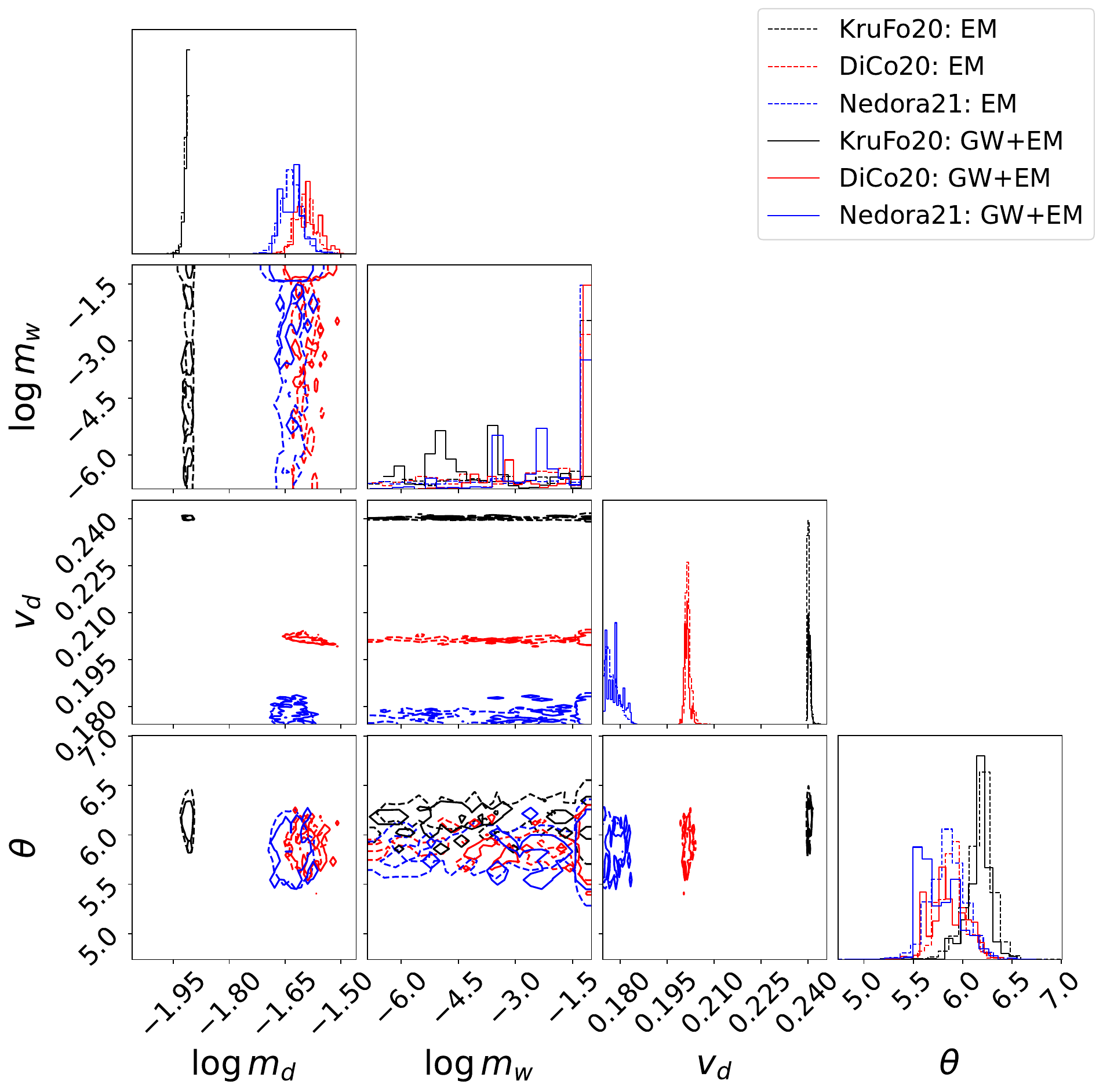}
    \includegraphics[width=\linewidth]{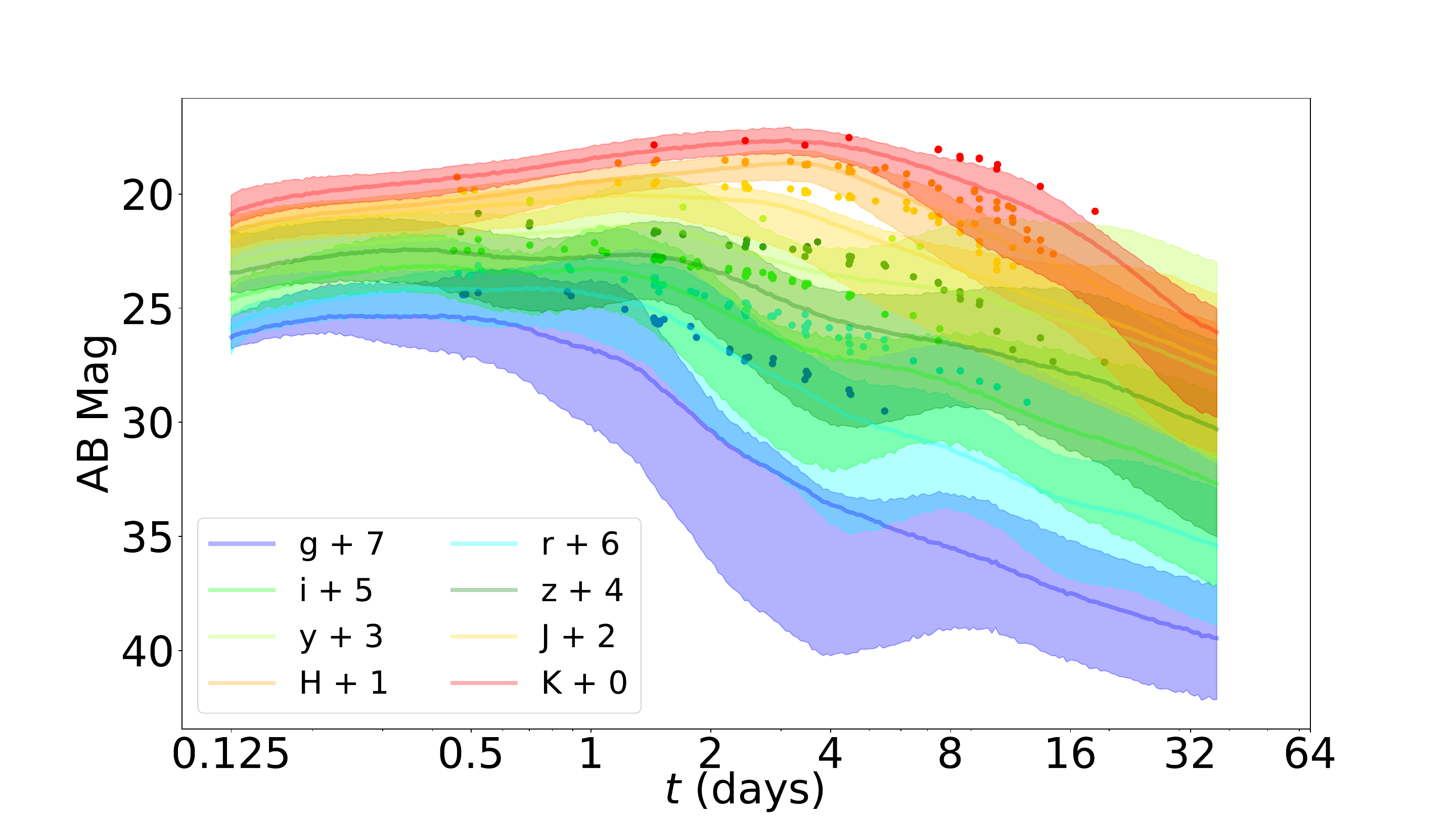}
    \caption{\textit{Top}: Ejecta posterior distributions using using EM (dashed) and GW+EM (solid) likelihoods with an assumed disk mass ejection of \fdisk$=30\%$. In contrast to Figure \ref{fig:corner_ejecta_narrow_mc_fix_vw}, $m_w$ is less constrained (leading to the large uncertainties in the light curves below) and $v_d$ is specific to the ejecta fit model.
    \textit{Bottom}: Light curves associated with the \Nedora\ GW+EM ejecta posteriors presented above, with solid lines corresponding to the median posterior values and shaded bands representing 1$\sigma$ uncertainty. It is obvious that, at least for the purposes of parameter inference, assuming the same fixed disk fraction for all three ejecta fits yields poor fits to the data with a relatively high degree of uncertainty.}
    \label{fig:corner_ejecta_fixed}
\end{figure}

We also consider the case of a fixed disk mass ejection fraction, in which \fdisk is not a free parameter, but restricted to the fixed value \fdisk $= 30\%$. 
The purpose of this analysis is to evaluate the behavior of the ejecta fits in predicting light curves under the assumption of a generally applied ejected disk fraction. 
The top half of Figure~\ref{fig:corner_ejecta_fixed} displays the same ejecta parameters as in Figure~\ref{fig:corner_ejecta_narrow_mc_fix_vw}. 
While the dynamical mass $m_{\rm{d}}$ and viewing angle $\theta$ are somewhat consistent with prior results, the wind mass $m_{\rm{w}}$ and dynamical velocity $v_{\rm{d}}$ posteriors yield substantially different results.
In the bottom half of Figure~\ref{fig:corner_ejecta_fixed}, we plot the light curves corresponding to the \Nedora\ GW+EM posteriors shown in the top part of Figure~\ref{fig:corner_ejecta_fixed}. 
While the light curves corresponding to the median posterior values (solid lines) are clearly worse fits than those in Figure~\ref{fig:lc_unconstrained}, the more striking feature lies in the exceptionally large uncertainties across all broadband filters.
More specifically, the uncertainty bounds are much broader than before, resulting from the broad dynamic range of the $m_{\rm{w}}$ posteriors. 
The difficulty in constraining $m_{\rm{w}}$ comes from \fdisk being directly tied to the amount of ejected wind mass; 
as such, what the $m_w$ posteriors in Figure~\ref{fig:corner_ejecta_fixed} are really showing us is the Monte Carlo integrator trying to reconcile the values of $m_w$ that actually fit the data with the constricting value of \fdisk $= 30\%$ that fixes the wind ejecta mass.

All of the samples generated by our analyses, including the GW likelihood data, can be found at \cite{GWEM_data_repo}.

\section{Discussion}
\label{sec:discussion}

We perform multimessenger inference of GW170817/AT2017gfo using three different ejecta models and a single kilonova model represented by the neural network surrogate provided in Ref.~\cite{Peng24}. 
This surrogate, when compared with observations, leads to strong constraints on our kilonova model's observable parameters, with posterior extent likely comparable to the information extractable from a perfect model.
While many systematic uncertainties have still yet to be incorporated into our analysis, many of which are well
described in a recent study \cite{2024ApJ...975..213B}, our surrogate model is nevertheless sufficient to assess the
prospects of, and challenges facing, multimessenger investigations.
Thus, this surrogate provides us with an ideal opportunity to assess the consistency of and perform systematic uncertainty quantification of our methods to perform multimessenger inference, focusing on the remaining ingredients of that challenge.  
We find that the three ejecta fits we consider largely prefer asymmetric mass binaries (though one allows for symmetry), with all three ejecta fits encompassing effectively different regimes of the neutron star radius $R_{\rm{1.4}}$.

Unsurprisingly, inference analyses using a joint EM and GW dataset are more constraining than using EM-only data. 
This behavior is most evident for the $\mathcal{M}_c$ and $\delta$ posteriors in Figure~\ref{fig:corner_binary_narrow_mc_fix_vw}, which highlight the impact of including the gravitational likelihood $\mathcal{L}_{\rm{GW}}$. 
The $\mathcal{M}_c$ posteriors for each fit clearly converge on a preferred value when supplemented with GW information; 
likewise, the $\delta$ posteriors become narrower on both global (dynamic range of posterior values) and local (per-ejecta-fit posterior values) scales.
With regard to the NS radius $R_{\rm{1.4}}$ posteriors, the difference between EM and GW+EM analyses is less striking.
More prevalent, however, are the different regions of the NS radius parameter space covered by the posteriors corresponding to each ejecta fit.
Specifically, the GW+EM predictions for $R_{\rm{1.4}}$ are roughly between 8--15~km, 10--20~km, and 10--40~km for the \Nedora\, \KruFo\, and \DiCo\ forward models, respectively.
These results strongly imply that, provided a constraining kilonova model, multimessenger inference about neutron star radius and, consequently, the dense matter equation of state is entirely dependent on the chosen forward model \editedtwo{and, more importantly, the chosen prescription for mapping BNS merger ejecta to kilonova model input ejecta}.

Keeping \edited{in mind the individual contributions from the GW and EM information}, the joint GW/EM inferences shown in Figure~\ref{fig:corner_binary_narrow_mc_fix_vw} make sense. For almost all parameters, the addition of GW information only
slightly perturbs the answers derived from EM information alone.  
This behavior is most evident in the bottom panel of Figure~\ref{fig:corner_binary_narrow_mc_fix_vw}, where narrower priors on \fdisk directly affect the ejected wind mass and thus the EM likelihood, shifting the $\delta$ and $R_{\rm{1.4}}$ posteriors to better agreement across ejecta fits.

As explained in Section~\ref{sec:kn_ejecta_fits}, we introduce the $\alpha_{\rm{dyn}}$, $f_{\rm{disk}}$, and $\beta_\phi$ 
free parameters to quantify the systematic uncertainty of the ejecta fit predictions, 
with \sigmasys encompassing all other sources of systematic uncertainty in our light curves.
As previously mentioned, we do not plot \sigmasys in Figure~\ref{fig:corner_binary_narrow_mc_fix_vw} because of its largely unconstrained and uninformative behavior.
Conversely, Figure~\ref{fig:corner_binary_narrow_mc_fix_vw} shows that \fdisk is the most constrained of these parameters, 
while the \alphadyn posteriors are generally broader and show minimal preference across individual ejecta fits.
The velocity systematics $\beta_\phi$ lie somewhere in between, with the Nedora and KruFo GW+EM analyses preferring slower velocities ($\beta_\phi > 0$)
 and the KruFo EM and DiCo analyses preferring faster velocities ($\beta_\phi < 0$).
For the velocity values allowed by the range of $\beta_\phi$, we find that, compared to the recovered value of $v_{\rm{ej}} = v_{\rm{d}}$ = 0.14 from Figure~\ref{fig:corner_ejecta_narrow_mc_fix_vw}, the \KruFo, \DiCo, and \Nedora\ forward models recover relative uncertainties on the velocity of 30.7\%, 138\%, and 63\%. 
For all but the \KruFo\ fit, these relative differences are substantially higher than the 33\% originally reported in Ref.~\cite{2017CQGra..34j5014D} from which the ejecta velocity fits originate.

Given the narrow peaks of the \fdisk posteriors, we interpret the \fdisk parameter as setting the scale for the wind ejecta mass, 
with the $\alpha_{\rm{dyn}}$, $\beta_\phi$, and \sigmasys parameters allowing our inference enough flexibility to fit the AT2017gfo light curves. 
Likewise, we expect the ejecta mass scale set by \fdisk to be the dominant parameter determining the light curve behavior (see, e.g., \cite{2023PhRvR...5d3106R}); however, recent studies have shown that velocity profiles can contribute significantly to variations in the light curves \cite{2023ApJ...958..121T, 2024ApJ...961....9F}.
Our velocity systematics can have substantial effects on inference, corroborating the findings of these recent studies.
These results motivate more detailed studies of versatile ejecta fits which, among other effects, specifically focus
on a separate treatment of dynamical and secular ejecta velocities.

Within the context of no velocity systematics ($\beta_\phi = 0)$, a closer look at the $R_{\rm{1.4}}$ vs. $\delta$ two-dimensional posterior in Figure~\ref{fig:corner_binary_narrow_mc_fix_vw_nobf} indicates a cubic relationship between the two parameters. 
Specifically, as the mass asymmetry of the binary increases $\delta \rightarrow 0.5$, the radius of the individual stars decreases $R_{\rm{1.4}} \rightarrow 0$, indicating that more asymmetric binaries prefer a softer (more compressible) EOS.
This behavior is exactly the same as that observed in Figure~\ref{fig:contours}, where the ejecta velocity contours (dotted lines ranging from 0.1 to 0.2) dictated the $R_{\rm{1.4}}$ vs. $\delta$ relationship.
The persistence of this restrictive behavior from the naive contours in Figure~\ref{fig:contours} to our posteriors in Figure~\ref{fig:corner_binary_narrow_mc_fix_vw_nobf} strongly suggests that future analyses of the NS radius and EOS must include \textit{both} mass and velocity systematic parameters, lest they introduce significant biases on top of the ones already present in selecting an ejecta forward model.

Finally, in the case where the fraction of ejected disk material is fixed to \fdisk $= 30\%$, we recover low quality posteriors that clearly struggle to fit the data. 
Specifically, the wind mass $m_{\rm{w}}$ is unconstrained compared to the dynamical mass $m_{\rm{d}}$, resulting in prohibitively large uncertainties in the light curves.

\section{Conclusion}
\label{sec:conclusion}

In this paper, we have examined multimessenger Bayesian inference of GW170817 and its kilonova AT2017gfo using
forward-modeling ejecta predictions which are based on numerical relativity simulations of neutron star mergers.
Using a fixed baseline kilonova model \cite{Peng24} \editedtwo{and an established, yet imperfect mapping between BNS merger and kilonova ejecta}, we assess the impact of systematic uncertainty in ejecta modeling,
using three forward model ejecta fits presented in Refs. \cite{2020PhRvD.101j3002K, 2020Sci...370.1450D,
  2022CQGra..39a5008N}.    We find that these three forward models are in tension with one another and necessarily
over-constrain ejecta inference (i.e., they make more predictions than they have parameters).   Using concrete
interpretation of GW170817/AT2017gfo, and after adding a generous budget for systematic uncertainty above and beyond
stated confidence, we illustrate the impact of the tension in these three models on multimessenger
interpretation: we find distinctly different conclusions about the underlying binary with each ejecta model, even
allowing for systematics.   We furthermore point out that adopting stronger assumptions about ejecta, which omit one or more of
our systematic uncertainty parameters, lead to even more extreme conclusions about AT2017gfo/GW170817. \edited{The posteriors we infer when omitting ejecta velocity systematics directly challenge the results from previous multimessenger analyses which marginalize over the ejecta velocity entirely \cite{2019MNRAS.489L..91C, 2020Sci...370.1450D}.}

Our study builds upon and extends the work in Ref. \cite{Peng24}, which emphasized purely phenomenological kilonova inference with a fiducial
$\sigma_{\rm sys}=0.5$ or a similar prior on $\sigma_{\rm sys}$, without ejecta forward modeling.  Except for when we
strongly limit our assumptions about ejecta systematics, our inferences about the observed kilonova ejecta are identical. We highlight that key
elements of these ejecta fits may be used well outside of their region of calibration (e.g., ejecta velocity) when
used in conjunction with the surrogate kilonova model from our prior work in Ref. \cite{Peng24}.

The primary observations highlighted by our analysis are:

\begin{itemize}
    \item Fits to numerical relativity simulations of neutron star mergers which predict mass ejection suffer from
      systematic uncertainties stemming from the parameter coverage and accuracy of  the simulations used to derive them.
Unfortunately, multimessenger inference requires confidence in \editedtwo{the imperfect mapping between BNS merger and kilonova model ejecta, guiding} these predictions well outside conventional
configurations to properly assess extreme scenarios. \edited{At present, we are unable to confidently and consistently connect binary and ejecta parameters using GW170817/AT2017gfo's rich observational datasets, thus also dooming future multimessenger inferences to a similar fate.}
      
    \item Given a kilonova model which can provide tight constraints on ejecta parameters, multimessenger inference
      about binary parameters like the neutron star radius and, consequently, the dense matter equation of state are
      strongly dependent on the chosen forward model and its ejecta predictions \editedtwo{which are primarily guided by the chosen prescription for mapping BNS merger ejecta to kilonova model input ejecta}.
      The ejecta velocity in particular
      seems to have a strong impact, as ours and other contemporary inferences favor velocities at the low end of
      calibrated fits. 
      
    \item \edited{Our study examines the inconsistencies between ejecta fits introduced by mapping binary parameters to ejecta mass \emph{and} velocity, challenging the results from previous multimessenger studies which marginalize over ejecta velocity.} Corroborating previous results (e.g. \cite{2023ApJ...958..121T, 2024ApJ...961....9F}), we find that a
      treatment of \textit{both} mass and velocity systematic uncertainties is required for maximally unbiased inference
      of ejecta forward model predictions.
      
    \item Within the context of our model, we find that the fraction of the ejected disk lies around $4-16\%$ for a
      GW170817-like event, with said fraction also dependent on the chosen forward model. This agrees with predictions
      from sub-second GRMHD simulations with full neutrino transport, but is in tension with assumptions typically made in the
      literature (up to $40\%$ of disk ejection).

      \item \editedtwo{We emphasize that some of the outcomes and discussed pitfalls in this work are likely to also be related to the specific, established choice of mapping between BNS merger ejecta and kilonova model ejecta inputs. Although conventions exist in the literature, they are incomplete, not one-to-one, and additional work establishing more physical mappings is necessary for accurate multimessenger inference.}
\end{itemize}

Our results suggest that there are currently no broadly applicable prescriptions which accurately map binary parameters 
to ejecta predictions or correctly assume the ejected fraction of an arbitrary post-merger accretion disk.
\edited{A large portion of the mapping incompatibility boils down to the difficulty in prescribing the ejecta from numerical relativity simulations to the standard assumption of two distinct, angle-separated kilonova ejecta components.}
The lack of such prescriptions strongly motivates future studies which examine the connection between 
numerical relativity simulations of neutron star mergers and their ejecta predictions.
While future multimessenger observations of neutron star mergers will shed more light on the broader population of mergers, in the meantime, 
our conclusions suggest that significantly more additional numerical relativity studies are necessary to understand the connection between binary and ejecta parameters.
Additionally, for kilonova models to be confidently used in multimessenger analyses such as the proof of concept presented in this manuscript, more work is required to improve the fidelity of the radiative transfer simulations used to model these events.
Likewise, in understanding these connections, we encourage future studies to carefully consider the contexts within
which each of the ejecta forward models is applicable and assume appropriate levels of uncertainty in their analyses.

\section{Acknowledgments}
ROS and MR acknowledge support from NSF AST 1909534 and 2206321. KW acknowledges support from NSF AST-2319326 and
PHY-2012057.  ROS also acknowledges support via NSF PHY 2012057, 2309172, and
the Simons Foundation. The work by CLF, CJF, MR, MRM, OK, and RTW was supported by the US Department of Energy through the Los Alamos National Laboratory (LANL). This research was made possible in part by the Information Science \& Technology Institute (ISTI) at LANL. This research used resources provided by LANL through the institutional computing program. Los Alamos National Laboratory is operated by Triad National Security, LLC, for the National Nuclear Security Administration of U.S.\ Department of Energy (Contract No.\ 89233218CNA000001). This document has been assigned LA-UR-25-22248.

\appendix*

\section{Interpreting the \DiCo\ \texorpdfstring{$\tilde{\Lambda}$}{Lt} Results}

\begin{figure}
    \centering
    \includegraphics[width=\linewidth]{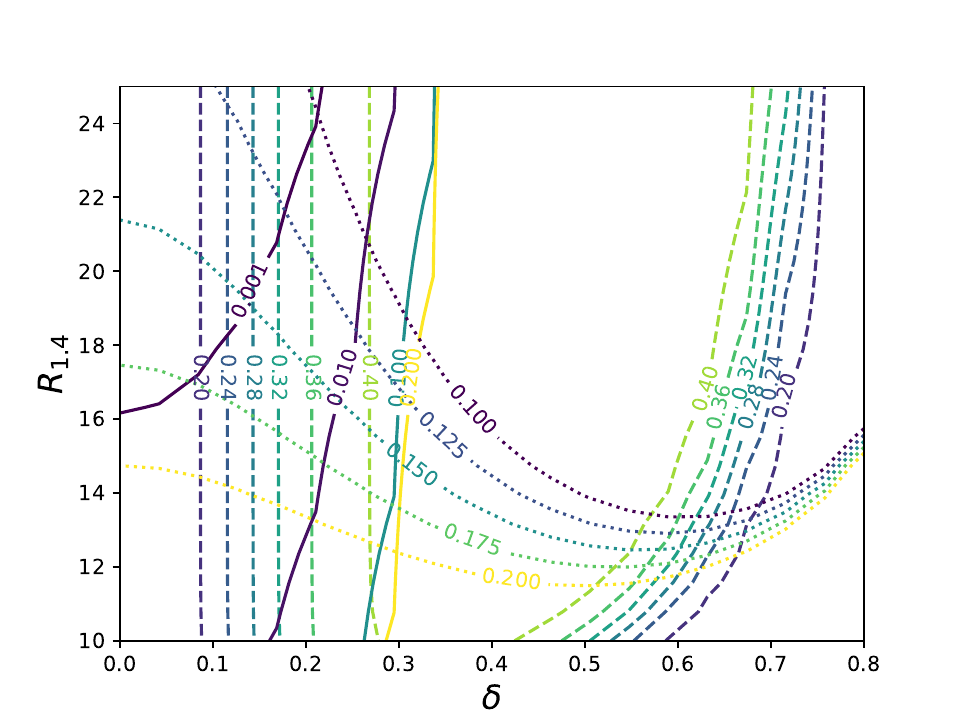}
    \caption{Ejecta contours as in Figure~\ref{fig:contours}, but for the \DiCo\ forward model. The disk mass (dashed lines) is constrained by $\delta$ but unbounded in the $R_{\rm{1.4}}$ dimension.}
    \label{fig:contours_DiCo}
\end{figure}

The \DiCo\ $\tilde{\Lambda}$ and $R_{\rm{1.4}}$ posteriors in Figures~\ref{fig:GW_parameter_posterior} and \ref{fig:corner_binary_narrow_mc_fix_vw} are appreciably different from those corresponding to the \KruFo\ and \Nedora\ ejecta fits, favoring much larger radii and $\tilde{\Lambda}$ than informed by the GW likelihood.
We examine the \DiCo\ ejecta contours in Figure~\ref{fig:contours_DiCo}.
The velocity prescription remains the same, while the disk mass and dynamical ejecta fits change. The dynamical ejecta contours behave similarly to the \KruFo\ contours, with larger values of $\delta$ required to produce a higher quantity of dynamical ejecta.
The disk mass contours (dashed lines) are the most different between the two fits.
Whereas in the \KruFo\ fits, disk mass was weakly dependent on $\delta$ and primarily set by $R_{\rm{1.4}}$, the \DiCo\ fits behave oppositely, with disk mass defined by $\delta$ and completely agnostic to $R_{\rm{1.4}}$.
This reversed dependency enables the \DiCo\ posteriors to find solutions near $\delta=0$ with low $\chi_{\rm{eff}}$, in line with the GW-informed likelihood; however, the lack of a $R_{\rm{1.4}}$ constraint results in the uncharacteristically large $\tilde{\Lambda}$ values observed in Figure~\ref{fig:GW_parameter_posterior}.

\bibliography{bibliography,gw-astronomy-mergers-ns-nuclearphysics,gw-astronomy-mergers,gw-astronomy-mergers-ns-gw170817,gw-astronomy-mergers-approximations,LIGO-publications}

\end{document}

%% file: _preamble.tex
\usepackage{graphicx}
\usepackage{bm}
\usepackage{mathtools}
\usepackage{hyperref}
\usepackage{amsmath}
\usepackage{mathrsfs}
\usepackage[dvipsnames]{xcolor}
\usepackage{soul}
\usepackage{orcidlink}
\usepackage{hyperref}
\hypersetup{
 pdftitle={GWEM Paper}, % title
 pdfauthor={Risti\'c}, % author
 colorlinks=true,
 citecolor={blue},
}

\newcommand\unit[1]{{\rm #1}}

\newcommand\KruFo{KruFo20}
\newcommand\DiCo{DiCo20}
\newcommand\Nedora{Nedora21}
\newcommand\alphadyn{$\alpha_{\rm{dyn}}$ }
\newcommand\fdisk{$f_{\rm{disk}}$ }
\newcommand\sigmasys{$\sigma_{\rm{sys}}$ }

\newcommand\edited[1]{{\color{black}#1}}
\newcommand\editedtwo[1]{#1}

\newcommand{\apjl}{Astrophysical Journal Letters}
\newcommand{\apjs}{Astroph. J. S.}

\newcommand{\pasj}{PASJ}

\newcommand{\pasa}{PASA}
\newcommand{\mnras}{Monthly Notices of the Royal Astronomical Society} 
\newcommand{\aap}{Astronomy and Astrophysics}

\newcommand{\prx}{Physical Review X}

\def\RIT{Center for Computational Relativity and Gravitation, Rochester Institute of Technology, Rochester, New York
 14623, USA}
\def\XCP{Computational Physics Division, Los Alamos National Laboratory, Los Alamos, NM, 87545, USA}
\def\CTA{Center for Theoretical Astrophysics, Los Alamos National Laboratory, Los Alamos, NM 87545, USA}
\def\CCSS{Computer, Computational, and Statistical Sciences Division, Los Alamos National Laboratory, Los Alamos, NM
 87545, USA}
\def\TD{Theoretical Division, Los Alamos National Laboratory, Los Alamos, NM 87545, USA}
\def\UA{The University of Arizona, Tucson, AZ 85721, USA}
\def\NM{Department of Physics and Astronomy, The University of New Mexico, Albuquerque, NM 87131, USA}

\def\GWU{The George Washington University, Washington, DC 20052, USA}

\graphicspath{{figures/}}